 \documentclass{aa}  
\usepackage{graphicx}
\usepackage{txfonts}
 \newcommand{ \PD    }[2]{ \frac{ \partial #1 }{ \partial #2 } }

\begin{document} 

\title{The Formation and Destruction of Molecular Clouds and 
       Galactic Star Formation}

\subtitle{An Origin for The Cloud Mass Function 
          and Star Formation Efficiency
          }

\titlerunning{Galactic Star Formation}
\authorrunning{Inutsuka, Inoue, Iwasaki, Hosokawa}

\author{Shu-ichiro Inutsuka \inst{1}, 
        Tsuyoshi Inoue,  \inst{2}, 
        Kazunari Iwasaki \inst{1,3},
        \and
        Takashi Hosokawa \inst{4}
        }

\institute{
 Department of Physics, Graduate School of Science, Nagoya University\\
 Nagoya 464-8602, Japan
    \email{inutsuka@nagoya-u.jp}
 \and
 Division of Theoretical Astronomy, National Astronomical Observatory of Japan\\ 
 Osawa, Mitaka, Tokyo 181-8588, Japan
 \and
 Department of Environmental Systems Science, Doshisha University\\ 
 Tatara Miyakodani 1-3, Kyotanabe City, Kyoto 610-0394, Japan
 \and
 Department of Physics and Research Center for the Early Universe\\
 The University of Tokyo, Tokyo 113-0033, Japan 
}

\date{Received December 24, 2014; accepted May 18, 2015}

\abstract{
We describe an overall picture of galactic-scale star formation. 
Recent high-resolution magneto-hydrodynamical simulations of two-fluid dynamics with cooling/heating and thermal conduction have shown that the formation of molecular clouds requires multiple episodes of supersonic compression.  
This finding enables us to create a scenario in which molecular clouds form in interacting shells or bubbles on a galactic scale. 
First we estimate the ensemble-averaged growth rate of molecular clouds over a timescale larger than a million years. 
Next we perform radiation hydrodynamics simulations to evaluate the destruction rate of magnetized molecular clouds by the stellar FUV radiation. 
We also investigate the resultant star formation efficiency within a cloud which amounts to a low value (a few percent) if we adopt the power-law exponent $\sim -2.5$ for the mass distribution of stars in the cloud. 
We finally describe the time evolution of the mass function of molecular clouds over a long timescale (>1Myr) and discuss the steady state exponent of the power-law slope in various environments. 
}
 
\maketitle
 
\section{Introduction}
Recent observational studies of nearby star-forming regions with the {\em Herschel Space Observatory} have convincingly shown that stars are born in self-gravitating filaments 
 \citep[e.g., ][]{Andre+2010,Arzoumanian+2011}.  
In addition, the resultant mass function of star-forming dense cores are now explained by the mass distribution along filaments \citep{Inutsuka2001,Andre+2014}. 
This simplifies the question of the initial conditions of star formation, but poses the question of how such filamentary molecular clouds are created in the interstellar medium (ISM) prior to the star formation process. 
Recent high-resolution magneto-hydrodynamical simulations of two-fluid dynamics with cooling/heating and thermal conduction by \citet{InoueInutsuka2008,InoueInutsuka2009} have shown that the formation of molecular clouds requires multiple episodes of supersonic compression \citep[see also][]{Heitsch+2009}. 
\citet{InoueInutsuka2012} further investigated the formation of molecular clouds in the magnetized ISM 
and revealed the formation of a magnetized molecular cloud by the accretion of HI clouds created through thermal instability. 
Since the mean density of the initial multi-phase HI medium is an order of magnitude larger than the typical warm neutral medium (WNM) density, this formation timescale is shorter than that of molecular cloud formation solely through the accumulation of diffuse WNM 
\citep[see, e.g.,][for the cases of WNM flows]{KoyamaInutsuka2002,Hennebelle+2008,HeitschHartmann2008,Banerjee+2009,Vazquez-Semadeni+2011}. 
The resulting timescale of molecular cloud formation of $\gtrsim$10 Myrs is consistent with the evolutionary timescale of molecular clouds in the LMC \citep{Kawamura+2009}.

We have done numerical simulations of additional compression of already-formed but low-mass molecular clouds, and found interesting features associated with realistic evolution.
Figure 1 shows a snapshot of the face-on view of the layer created by compressing a non-uniform molecular cloud with a shock wave propagating at 10 km/s. The direction of the shock compression is perpendicular to the layer. The magnetic field lines are mainly in the dense layer of compressed gas. 
The strength of the initial magnetic field prior to the shock compression is $20\mu$Gauss and that of the dense region created after compression is about $200\mu$Gauss on average. 
Many dense filaments are created with axes perpendicular to the mean magnetic field lines. 
We can also see many faint filamentary structures that mimic ``striations'' observed in the Taurus Dark Cloud and are almost parallel to the mean magnetic field lines \citep[][]{Goldsmith+2008}. 
In our simulations, these faint filaments appear to be feeding gas onto dense filaments (similar to what is observed for local clouds by \citet[e.g.,][]{Sugitani+2011,Palmeirim+2013,Kirk+2013}). 
Once the line-mass of a dense filament exceeds the critical value ($2C_{\rm s}^2/G$), star formation is expected to start \citep{InutsukaMiyama1992,InutsukaMiyama1997,Andre+2010}. 
This threshold of line-mass for star formation is equivalent to the threshold of the column density of molecular gas $116M_\sun {\rm pc}^{-2}$ \citep[][]{Lada+2010}, if the widths of filaments are all close to 0.1pc \citep[][]{Arzoumanian+2011,Andre+2014}. 

Although further analysis is required for quantitative comparison between the results of simulation and observed structures, Figure 1 clearly shows that the structures created by multiple shock wave passages do match the characteristic structures observed in filamentary molecular clouds. 
This motivates us to describe a basic scenario of molecular cloud formation.
The present paper is focused on the implications of this identification of the mechanism of molecular cloud formation.

\begin{figure}
\centering
\includegraphics[width=\hsize]{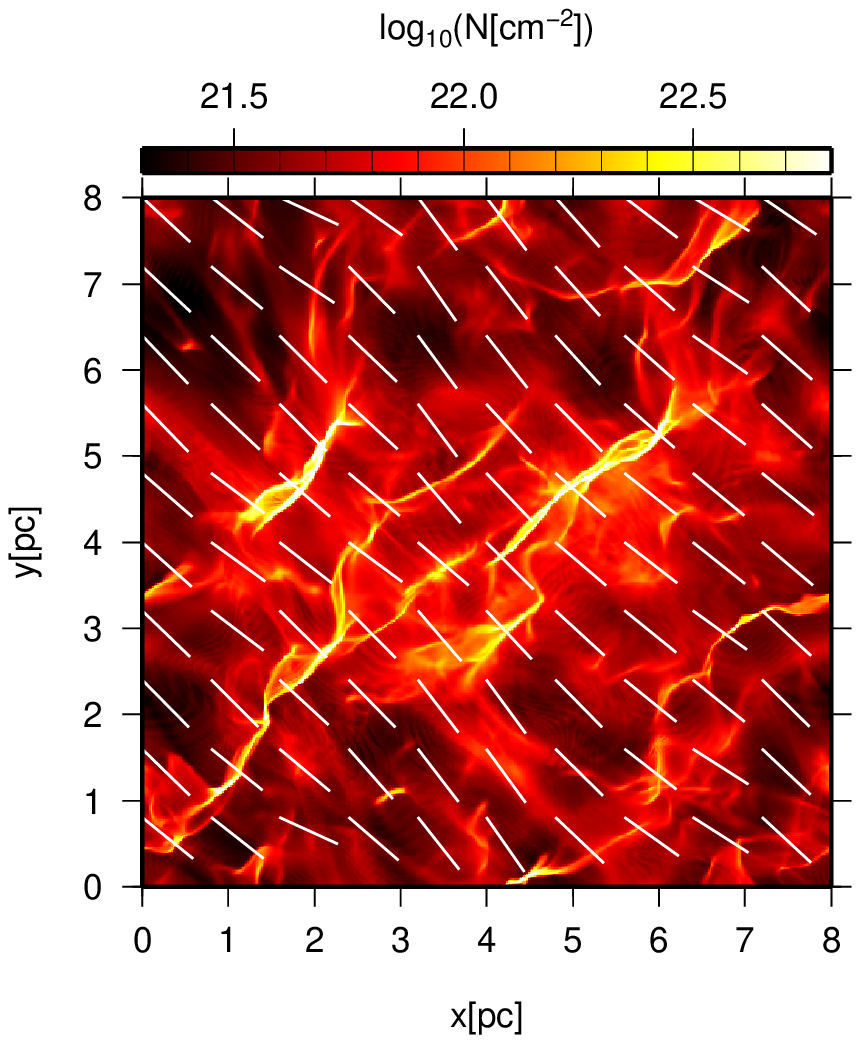}
\caption{Face-on column density view of a shock-compressed dense layer of molecular clouds.  
We set up low-mass molecular clouds by the compression of two-phase HI clouds. 
This snapshot shows the result of an additional compression of 
low-mass
molecular clouds by a shock wave propagating at 10 km/s. 
The magnetic field lines are mainly in a dense sheet of a compressed gas. 
The color scale for column density (in cm$^{-2}$) is shown on top. 
The mean magnetic field is in the plane of the layer and its direction is shown by white bars.
Note the formation of dense magnetized filaments whose axes are almost perpendicular to the mean magnetic field. 
Fainter ``striation''-like filaments can also be seen, that are almost perpendicular to the dense filaments. 
\label{fig1}}
\end{figure}

\section{A Scenario of Cloud Formation Driven by Expanding Bubbles}
\begin{figure}
\centering
\includegraphics[width=\hsize]{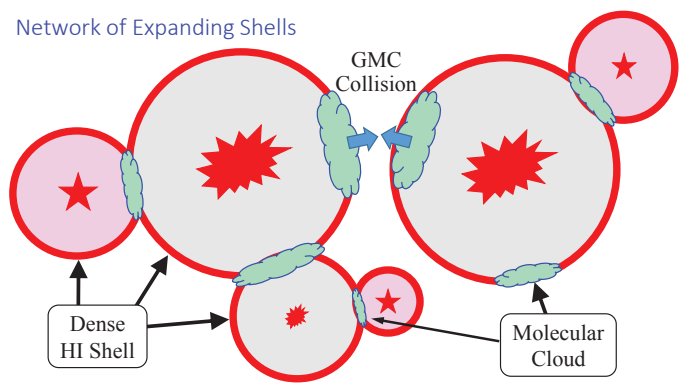}
\caption{
A schematic picture of sequential formation of molecular clouds by multiple compressions by overlapping dense shells driven by expanding bubbles. 
The thick red circles correspond to magnetized dense multi-phase ISM where cold turbulent HI clouds are embedded in WNM.  
Molecular clouds can be formed only in the limited regions where the compressional direction is almost parallel to the local mean magnetic field lines, or in regions experiencing an excessive number of compressions.  
An additional compression of a molecular cloud tends to create multiple filamentary molecular clouds. 
Once the line-mass of a filament exceeds the critical value, even in a less massive molecular cloud, star formation starts. 
In general, star formation in a cloud accelerates with the growth in total mass of the cloud. 
Giant molecular clouds collide with one another at limited frequency. 
This produces very unstable molecular gas and may trigger very active star formation.  
\label{fig2}}
\end{figure}

HI observations of our Galaxy reveal many shell-like structures near the galactic plane \citep[e.g.,][]{HartmannBurton1997,Taylor+2003}. 
We identify repeated interactions of expanding shock waves as a basic mechanism of molecular cloud formation, and we depict the overall scenario of cloud formation in our Galaxy as a schematic picture in Figure \ref{fig2}. 
In this picture, red circles correspond to the remnants of shock waves due to old and slow supernova remnants or  expanding HII regions. 
Cold HI clouds embedded in WNM are almost ubiquitously found in the shells of these remnants  
\citep[e.g.,][]{HartmannBurton1997,Taylor+2003,2007ApJ...664..363H}. 
Molecular clouds are expected to be formed in limited regions where the mean magnetic field is parallel to the direction of shock wave propagation, or in regions where an excessive number of shock wave sweepings are experienced. 
Therefore, molecular clouds can be found only in limited regions in shells.  
Note that the typical timescale of each shock wave is of order 1Myr, but the formation of molecular clouds requires many Myrs. 
Some bubbles become invisible as a supernova remnant or an HII region many million years after their birth. 
Therefore, this schematic picture corresponds to a ``very long exposure snapshot'' of the real structure of the ISM. 
Each molecular cloud may have random velocity depending on the location in the most recent bubble that interacts with the cloud. 
Interestingly, this multi-generation picture of the evolution of molecular clouds seems to agree with the observational findings of \citet{Dawson+2011a,Dawson+2011b,Dawson+2015}, who investigated the transition of atomic gas to molecular gas in the wall of Galactic supershells.
In the case of LMC, \citet{Dawson+2013} concluded that only $12\sim25$\% of the molecular mass can be apparently attributed to the formation due to presently visible shell activity.  
This may not be inconsistent with our scenario since Dawson et al (2013) only considered HI supergiant shells, whereas molecular clouds in our model can form at the interface of much smaller bubbles and shells (which observationally are more difficult to identify and characterize in the HI data) and the timescale for cloud forming shells to become invisible is much shorter than the growth timescale of molecular mass.

A typical velocity of the shock wave due to an expanding ionization-dissociation front is 10km/s, as shown by Hosokawa \& Inutsuka (2006a), since it is essentially determined by the sound speed of ionized gas ($\sim 10^4$K).  
Iwasaki et al. (2011b) have shown that if a molecular cloud is swept-up by shock wave of ~10km/s, it moves with a velocity slightly less than the shock speed.  
Thus, the mean velocity of each molecular cloud should be somewhat smaller than that of the most recent shock wave. 
When the shock velocity of a supernova remnant is much higher than 10km/s, the resultant interaction would result in the destruction of molecular clouds.  
Therefore, the cloud-to-cloud velocity dispersion of molecular clouds should be similar to 10km/s.
According to this acquisition mechanism of random velocity, the velocity of a cloud is not expected to depend strongly on its mass.  
In other words, random velocities of molecular clouds of different masses are not expected to be in equipartition of energy ($M \delta v^2/2=$const.).   
Observations by Stark \& Lee (2005) have shown that the random velocities of low-mass molecular clouds ($< 2 \times 10^5 M\sun$) only vary by a few, with no dependence on cloud mass. 
These observations are therefore more consistent with our picture than a model in which molecular clouds acquire their relative velocities via mutual gravitational interaction.

In limited circumstances, created molecular clouds collide with one another. 
This produces highly gravitationally unstable molecular gas and may trigger very active star formation \citep[e.g.,][]{Fukui+2014}.  
Inoue \& Fukui (2013) have done magnetohydrodynamical simulations of a cloud-cloud collision and argue that it may lead to active formation of massive stars \citep[see also][]{Vaidya+2013}. 
This mode of massive star formation is not, however, a prerequisite of our model.

\subsection{Formation Timescale of Molecular Clouds} 
Let's first model the growth of molecular clouds. 
\cite{InoueInutsuka2012} have shown that we need multiple episodes of compression of HI clouds to create molecular clouds. 
According to the standard picture of supernova-regulated ISM dynamics (e.g., McKee \& Ostriker 1977), the typical timescale between consecutive compressions by supernova remnants is about 1Myr.  
The total creation rate of expanding bubbles is larger than the occurrence rate of supernova explosions, since the former can also be created by radiation from massive stars less massive than supernova progenitors.  
Therefore, the actual timescale of compressions in ISM, $T_{\rm exp}$, should be somewhat smaller than 1Myr if it is averaged over the Galactic thin disk.  
Obviously the compression timescale is smaller in the spiral arms and larger in inter-arm regions since star formation activity is concentrated in the spiral arms.
Thus, we have to consider the time evolution of cloud mass for much longer than 1 Myr. 

Let us estimate the typical timescale of molecular cloud growth. 
\cite{InoueInutsuka2012} have shown that the angle between the converging flow direction and the average direction of the magnetic field should be less than a certain angle for molecular cloud formation.  
Although Inoue \& Inutsuka (2009) shows that this critical angle depends on the flow speed, we adopt a critical angle of 15 degrees (=0.26 radian) in the following discussion for simplicity. 
This value is not so different from the angle ($\sim 20$ degrees) for possible compression in the simpler one-dimensional model by \cite{HennebellePerault2000}. 
For simplicity we assume that magnetic field is uniform in the region we consider and the direction of compression is isotropic. 
The solid angle spanned by the possible directions of compression resulting in the formation of a molecular cloud is $0.26^2 \pi$.  
The anti-parallel directions are also possible.  
Therefore, the probability, $p$, of successfully forming a molecular cloud in a single compression can be estimated by the ratio of solid angle over which compressions lead to molecular cloud formation to the solid angle of the whole sphere, i.e., $p=2 \cdot 0.26^2 \pi/(4\pi)=0.034$.
Note that Figure 1 is not the snapshot just after the birth of molecular clouds, but the result of one additional compression of the molecular clouds in which the direction of compression is perpendicular to the mean direction of the magnetic field lines.
We also emphasize that since the formation of a GMC requires many episodes of compression, our model does not predict a strong correlation between the present-day magnetic field direction and the orientation of the GMC.

After each compression a cloud may slightly expand because of the reduced pressure of the ambient medium, which may result in the loss of diffuse components of cloud mass. 
Observationally the average column densities of molecular clouds do not seem to change very much and always appear to correspond to a visual extinction of several. 
This means that the mass of a cloud is proportional to its cross-section. 
Since the compressional formation of molecular material is expected to be proportional to the cross-section of the pre-existing cloud, we can model the rate of increase of molecular cloud mass as 
\begin{equation}
  \frac{dM}{dt} = \frac{M}{T_{\rm f}} ,  \label{eq:Eqformation}
\end{equation}
where $T_{\rm f}$ denotes the formation timescale.  
This equation shows that resultant mass of each molecular cloud grows exponentially with a long timescale $T_{\rm f}$ if we average in time over a few Myr.
If self-gravity increases the accumulation rate of mass into the molecular cloud, the right-hand side of Equation (1) may have a stronger dependence on mass.  
For example, the so-called ``gravitational focusing factor'' increases the cross section of coalescence by a factor proportional to the square of mass for the large mass limit.  
This will produce a significantly steeper slope of the cloud mass function (see Section 4).   
A linear dependence on mass in our formulation implicitly assumes that self-gravity of the whole molecular cloud does not significantly affect the cloud growth.

Based on our investigation of molecular cloud formation described above, 
we estimate the formation timescale as follows:
\begin{equation}
  T_{\rm f} = \frac{1}{p} \cdot T_{\rm exp}.  \label{eq:Tformation}
\end{equation}
The average value in spiral arm regions would be $T_{\rm f} \sim 10$ Myr, but can be factor of a few longer in the inter-arm regions. 
In reality, many repeated compressions with large angles between the flow direction and mean magnetic field lines gradually increase the dense parts of clouds, and hence contribute to the formation of molecular clouds over a long timescale. 
This may mean that the actual value of $T_{\rm f}$ is somewhat smaller than the estimate of Equation (2). 

Fukui et al. (2009) has shown that the clouds with masses (a few $\times 10^5M_\sun$) gain their mass at a rate 0.05 $M_\sun$/yr over a timescale 10Myr.  
This means that the mass of a cloud in their sample doubles in $\sim 10$Myr, which is consistent with our choice of $T_{\rm f}=10$Myr. 
Note, however, that Fukui et al. (2009) argued that the atomic gas accretion is driven by the self-gravity of a GMC, which is not included in the present modelling where we assume that gas accretion is essentially driven by the interaction with expanding bubbles.  
If the gravitational force is significant for the HI accretion onto GMC, it possibly enhances the growth rate of molecular cloud (i.e., smaller $T_{\rm f}$).  
Further quantitative studies of the effect of self-gravity on the accretion of gas onto a GMC remain to be done.  
In the present paper we neglect this effect and do not distinguish self-gravitationally bound and pressure-confined clouds, for simplicity.

In regions where the number density of molecular clouds is very large, cloud-cloud collision may contribute to the increase of cloud mass, and hence, may also affect the mass function of molecular clouds.  
The detailed modelling of cloud-cloud collision will be given in our forthcoming paper.  
Here we ignore the contribution of cloud-cloud collision to the change of mass function and simply use the constant value of $T_{\rm f}$.

\section{Quenching of Star Formation in Molecular Clouds}
Next we consider the destruction of molecular clouds to determine how the star formation is quenched. 
Dale et al. (2012, 2013) have done extensive three dimensional simulations of star cluster formation with ionization or stellar wind feedback and shown that the effects of photo-ionization and stellar winds are limited in quenching the star formation in massive molecular clouds \citep[see also][]{Walch+2012}. 
\citet{Diaz-Miller+1998} calculated the steady-state structures of HII regions and pointed out that the photodissociation of hydrogen molecules due to FUV photons is much more important than photoionization due to UV photons for the destruction of molecular clouds. 
\cite{2005ApJ...623..917H,2006ApJ...646..240H,2006ApJ...648L.131H,2007ApJ...664..363H} actually included photodissociation in the detailed radiation hydrodynamical calculations of an expanding HII region in a non-magnetized ISM (by resolving photodissociative line radiation), and found the limited effect of ionizing radiation and essentially confirmed the critical importance of FUV radiation for the ambient molecular cloud. 

\subsection{Expanding HII Regions in Magnetized ISM} 
In the case of non-magnetized molecular gas of density $10^2{\rm cm}^{-3}$ around a massive star larger than $\sim 20 M_\sun$, a large amount of gas ($\sim 3\times 10^4 M_\sun$) is photodissociated and re-processed into molecular material in the dense shell around the expanding HII region within 5 Myrs \citep{2006ApJ...648L.131H}. 
According to the series of papers by \citet{InoueInutsuka2008,InoueInutsuka2009}, however, 
the inclusion of magnetic field is expected to reduce the density of the swept-up shell substantially.   
Therefore the magnetic field should affect significantly the actual structure of compressed shell and subsequent star formation process \citep[c.f., 3D simulation by][]{Arthur+2011}. 

To quantitatively analyze the consequence, we have done numerical magnetohydrodynamics simulations of an expanding bubble due to UV and FUV photons from the massive star. 
The details of the method is the same as described in 
\cite{2006ApJ...646..240H,2006ApJ...648L.131H} except for the inclusion of the magnetic field. 
Since the calculation assumes spherical symmetry, we include only the $13\mu$Gauss magnetic field that is transverse to the radial direction as a simplification.  
The magnetic pressure due to transverse field is accounted for in the Riemann solver as in \cite{Sano+1999} \citep[see][]{SuzukiInutsuka2006,IwasakiInutsuka2011}. 

\begin{figure}
\centering
\includegraphics[width=\hsize]{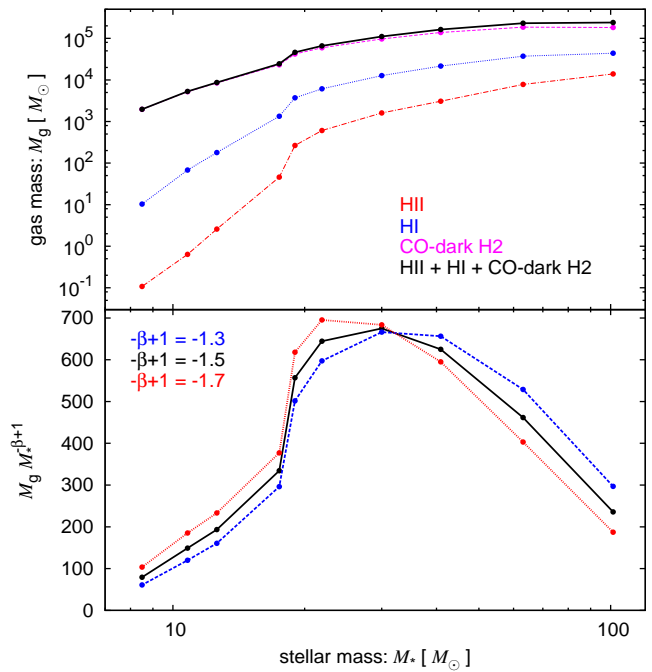}
\caption{
Upper Panel: 
Masses in various phases transformed from cold molecular gas around an expanding HII region at the termination time as a function of the mass of a central star. 
The red dot-dashed line, the blue dotted line, and purple dashed line correspond to 
ionized hydrogen in the HII region, 
neutral hydrogen in the photodissociation region, 
and warm molecular hydrogen gas without CO, respectively.  
The uppermost black solid line denotes the total mass of these non-star-forming gases. 
Lower Panel: 
The IMF-weighted mass of non-star-forming gas transformed from molecular gas by a massive star of mass $M_*$. 
The area under the curve is proportional to the mass generated by massive stars whose mass function follows $dn/d(\log M_*) \propto M_*^{-\beta +1}$. 
The peak of the curve determines the inverse of star formation efficiency $\epsilon_{\rm SF}$ (see explanation below Equation \ref{eq:SFE}). 
\label{fig3}
}
\end{figure}

The upper panel of Figure \ref{fig3} shows the resultant masses of ionized gas in HII region and atomic gas in photo-dissociation region transformed from cold molecular gas around an expanding HII region at the termination time as a function of the mass of a central star ($M_*$). 
Also plotted is the warm molecular gas in and outside the compressed shell around the HII region. 
The temperature of the warm molecular gas exceeds 50K. 
Its column density is smaller than $10^{21} {\rm cm}^{-2}$, and hence, dust shielding for CO molecule is not effective and all the CO molecules are photo-dissociated.
This warm molecular gas without CO (so-called, CO-dark H$_2$) is not expected to be gravitationally bound unless the mass of the parental molecular cloud is exceptionally large. 
Therefore the subsequent star formation in this warm molecular gas is not expected.  
The uppermost black solid line denotes the total mass ($M_{\rm g}(M_*)$) of these non-star-forming gases. 

The lower panel of Figure 3 shows gas mass in the upper panel multiplied by $M_*^{-1.3}$ (blue dashed curve), $M_*^{-1.5}$ (black solid curve), and $M_*^{-1.7}$ (red dotted curve).  
The areas under these curves are proportional to the mass affected by massive stars whose mass distribution follows 
$dn_*/d(log M_*) \propto M_*^{1.3}, M_*^{1.5}$, and $M_*^{1.7}$, respectively.  
We can see that the shape of the curve does not vary much, and stars with mass $20\sim30 M_\sun$ always dominate the disruption of the molecular cloud.

Our calculations include ionization, photodissociation, and magnetohydrodynamical shock waves, but are restricted to a spherical geometry.  
Therefore we should investigate the dispersal process in more realistic three dimensional simulations.  
However, the inclusion of photo-dissociation requires the numerical calculation of FUV line transfer and hence remains extremely difficult in multi-dimensional simulations.

\subsection{Star Formation Efficiency} 
Hereafter we assume the power law exponent of mass in the initial mass function of stars for large mass ($M_* > 8 M_\sun$) is $-\beta$ and $2<\beta<3$ 
($ dn_*/dM_* \propto M_*^{-\beta} $). 
Now we calculate the total mass of non-star forming gas disrupted by new born stars in a cloud. 
One might think that the total mass of non-star-forming mass in the stellar system can be calculated by 
$
   M_{\rm g,total} = \int_0^\infty M_{\rm g}(M_*) (dn_*/dM_*) dM_*
$.
However, this estimation is meaningful only in the case the number of massive stars in a cloud are very large, 
$
   \int_{20 M_\sun}^\infty (dn_*/dM_*) dM_* \gg 1. 
$ 
In reality, the number of massive stars in a molecular cloud of intermediate mass is quite small, and even a single massive star can destruct the whole parental molecular cloud. 
Thus, to analyze the quenching of star formation in molecular clouds, it is more appropriate to determine the most likely mass of the star that is responsible for the destruction of molecular clouds. 

Since we assume the large mass side of the stellar initial mass function can be approximated by the power law of the exponent $-\beta$, we can express the mass function in logarithmic mass of massive stars created in a cloud as 
\begin{equation}
     \frac{dn_*}{d\log M_*} = M_* \frac{dn_*}{dM_*} 
                         = N_* \left( \frac{M_*}{M_\sun} \right)^{-\beta+1}
                         ~~{\rm for}~M_* > 8 M_\sun
                                                  \label{eq:IMF} 
\end{equation}
Note that the pre-factor $N_*$ is defined for the mass distribution of stars in the individual cloud we are analyzing. 
For convenience, we define the effective minimum mass ($M_{\rm *m}$) of a star in the hypothetical power law mass function by the following formula for the total mass in the cloud,  
\begin{eqnarray}
     M_{\rm *,total} 
& = & 
       \int_{0}        ^{\infty} M_* \frac{dn_*}{dM_*} dM_* \nonumber \\ 
& \equiv & 
       \int_{M_{\rm *m}}^{\infty} 
          N_* \left( \frac{M_*}{M_\sun} \right)^{-\beta+1} dM
     = \left( \frac{N_*}{\beta-2}     \right)
       \left( \frac{M_\sun}{M_{\rm *m}}\right)^{\beta-2} .
                                                          \label{eq:Matotal}
\end{eqnarray}

Suppose that a single massive star more massive than $M_{\rm *d}$ is created in the molecular cloud. 
This condition can be expressed as 
\begin{equation}
    1 =\int_{M_{\rm *d}}^{\infty} \frac{dn_*}{dM_*} dM_* = 
       \left( \frac{N_*}{\beta-1} \right)
       \left( \frac{M_\sun}{M_{\rm *d}} \right)^{\beta-1}   
                         ~~{\rm for}~M_* > 8 M_\sun .
                                                             \label{eq:MdNa}
\end{equation}
This equation relates $M_{\rm *d}$ and $N_*$. 
We can express the total mass of stars in the cloud as a function of $M_{\rm *d}$ by eliminating $N_*$ in equations (\ref{eq:Matotal}) and (\ref{eq:MdNa}), 
\begin{equation}
     M_{\rm *,total} =
       \left( \frac{\beta-1  }{\beta-2  } \right)
       \left( \frac{M_\sun   }{M_{\rm *m}} \right)^{\beta-2}  
       \left( \frac{M_{\rm *d}}{M_\sun   } \right)^{\beta-1}   
                         ~~{\rm for}~M_* > 8 M_\sun .
                                                             \label{eq:MtMd}
\end{equation}
Thus, $M_{\rm *,total} \propto M_{\rm *d}^{\beta-1}$ for $M_* > 8 M_\sun$.
Now we suppose that a molecular cloud of mass $M_{\rm cl}$ is eventually destroyed by UV and FUV photons from a star of mass $M_{\rm *d}$ born in the cloud, and hence, star formation in the cloud is quenched. 
The condition for this to occur can be written as 
$
    M_{\rm cl} = M_{\rm g}
$
 and 
$
    \epsilon_{\rm SF} M_{\rm cl} = M_{\rm *,total},  
$
where $\epsilon_{\rm SF}$ is the star formation efficiency (the ratio of the total mass of stars to the mass of the parental cloud).

If $\epsilon_{\rm SF}$ is smaller than the value that would satisfy the above condition, the cloud destruction is not sufficient and star formation continues using the remaining cold molecular material in the cloud, which in turn increases $\epsilon_{\rm SF}$. 
Thus, we expect that the actual evolution of a molecular cloud finally satisifies the above condition when the star formation is eventually quenched.
This means that the star formation efficiency should be given by 
\begin{equation}
  \epsilon_{\rm SF} = \frac{M_{\rm *,total}}{M_{\rm g}(M_{\rm *d})} 
= \left( \frac{\beta-1   }{\beta-2   } \right)
  \left( \frac{M_\sun    }{M_{\rm *m}} \right)^{\beta-2}
  \left( \frac{M_{\rm *d}}{M_\sun    } \right)^{\beta-1}
  \left( \frac{M_{\rm g} }{M_\sun    } \right)^{-1}.
                        \label{eq:SFE} 
\end{equation}
The preceding argument suggests that the value of star formation efficiency should take the minimum value of the right hand side of this equation,  
i.e., the maximum value of $M_{\rm g} M_{\rm *}^{-\beta+1}$ where  $M_{\rm g}$ is a function of $M_*$. 
Figure 3 shows that the maximum value of $M_{\rm g} M_{\rm *}^{-\beta+1}$ is attained at $M_* \sim 30M_\sun$ where $M_{\rm g}$ is about $10^5 M_\sun$.  
Therefore we can conclude that once a massive star of $M_{\rm *d} = 30 \pm 10 M_\sun$ is created, the star formation is eventually quenched in a cloud of mass $M_{\rm g}(M_{\rm *d}) \sim 10^5 M_\sun$.  
This corresponds to $\epsilon_{\rm SF} \sim 10^{-2}$, if we adopt $\beta=2.5$ and $M_{\rm *m} = 0.1M_\sun$.
The dependence of $\epsilon_{\rm SF}$ on $M_{\rm *m}$ is quite weak ($\beta-2 \sim 0.5$) as shown in Equation (7).   
It is also not sensitive on $\beta$ in the limited range ($2.3 < \beta < 2.7$) as shown in Figure 3.  
Thus, the authors think that this value of $\epsilon_{\rm SF} \sim 10^{-2}$ is  robust in typical star forming regions in our Galaxy.  
This argument may explain the reason for the low star formation efficiency in molecular clouds observationally found many decades ago \citep[e.g.,][]{ZuckermanEvans1974}. 

Note that a sharp increase of $M_{\rm g} M_*^{-\beta+1}$ is due to the sharp increase of UV/FUV luminosity at $M_* \sim 20 M_\sun$. 
Therefore a star much smaller than $20M_\sun$ is not expected to be the main disrupter of the molecular cloud.  
For example, the upper panel of Figure 3 shows that a $10M_\sun$ star can quench $\sim 10^3 M_\sun$ of the surrounding molecular material. 
However a $10^3 M_\sun$ molecular cloud is not likely to produce a $10M_\sun$ star unless $\epsilon_{\rm SF} \sim 1$ as can be seen in equation (\ref{eq:MtMd}), and hence, the destruction of $10^3 M_\sun$ cloud by a $10M_\sun$ star is not expected, in general.

If the initial mass function does not depend on the parent cloud mass as we assume here, the star formation efficiency is not expected to depend on mass for a cloud larger than $\sim 10^5 M_\sun$. 
This can be understood as follows. 
The number of UV/FUV photons is proportional to the number of massive stars, which increases with the mass of the cloud.  
However, the required number of photons also increases with the mass of the cloud. 
For example, a $10^6 M_\sun$ cloud will produce 10 stars with mass $> 30M_\sun$ 
if $\epsilon_{\rm SF} = 10^{-2}$, $\beta=2.5$, and $M_{\rm *m} = 0.1M_\sun$.
Then, these 10 massive stars will destroy $10 \times 10^5 M_\sun$ molecular gas.  
Therefore star formation in the whole molecular cloud is quenched when $\epsilon_{\rm SF} = 10^{-2}$. 
Therefore we can conclude that the star formation efficiency does not depend on the mass of the cloud if the shape of the initial mass function does not depend on the mass of the cloud. 

Now we can estimate the timescale for the destruction of a molecular cloud. 
Our calculation of the expanding ionization/dissociation front in the magnetized molecular cloud shows that a $\sim 10^5 M_\sun$ molecular cloud can be destroyed within 4 Myrs.
The actual destruction timescale, $T_{\rm d}$, should be the sum of the timescale of formation of a massive star and expansion timescale of the HII region, i.e., $T_{\rm d} \approx T_* + 4$Myr, where $T_*$ denotes the timescale for a massive star to form once the cloud is created. 
After one cycle of molecular cloud destruction over a timescale $T_{\rm d}$, only a fraction, $\epsilon_{\rm SF}$, of the molecular gas is transformed into stars. 
Therefore, the timescale to completely transform a molecular cloud to stars is $T_{\rm d}/\epsilon_{\rm SF} \sim 1.4$Gyr for $T_* \sim 10$Myrs and $T_{\rm d} \sim 14$Myrs. 
This may explain the so-called ``depletion timescale'' of molecular clouds that manifests observationally in the Schmidt-Kennicutt Law \citep[e.g.,][] {Bigiel+2011,Lada+2012,KennicuttEvans2012}. 

\section{Mass Function of Molecular Clouds} 
In order to describe the time evolution of the mass function of molecular clouds, $n_{\rm cl}(M)=dN_{\rm cl}/dM$, over a timescale much longer than 1 Myr, 
 we adopt coarse-graining of short-timescale growth and destruction of clouds, and describe the continuity equation of molecular clouds in mass space as 
\begin{equation}
  \PD{n_{\rm cl}}{t} + \PD{}{M} \left( n_{\rm cl} \frac{dM}{dt} \right)
               =  - \frac{n_{\rm cl}}{T_{\rm d}} ,  
\end{equation}
where 
$n_{\rm cl}(dM/dt)$ denotes the flux of mass function in mass space, 
$dM/dt$ describes the growth rate of the molecular cloud as given in Equation (1).  
The sink term on the right hand side of this equation corresponds to the destruction rate of molecular clouds in the sense of ensemble average.  
If the dynamical effects such as shear and tidal stresses contribute to the cloud destruction \citep[e.g.,][]{Koda+2009,DobbsPringle2013}, we should modify $T_{\rm d}$ in this equation. 
Since the left hand side of this equation should be regarded as the ensemble average, the term $1/T_{\rm d}$ represents the sum of the destruction rate of all the possible processes. 
Here we simply assume that the resultant $T_{\rm d}$ is not very different from our estimate of destruction due to radiation feedback from massive stars. 

According to the series of our work on the formation of molecular clouds, the molecular cloud as a whole is not necessarily created as a self-gravitationally bound object.  
Therefore, our modelling of the mass function of molecular clouds is not restricted to the self-gravitationally bound clouds.
However, our modelling is not intended to describe the spatially extended diffuse molecular clouds much larger than the typical size of the bubbles ($\lesssim100$pc).

A steady state solution of the above equation is 
\begin{equation}
  n_{\rm cl}(M) = \frac{N_0}{M_{\sun}} \left( \frac{M}{M_{\sun}} \right)^{-\alpha}, 
\end{equation}
where $N_0$ is a constant and 
\begin{equation}
  \alpha = 1 + \frac{T_{\rm f}}{T_{\rm d}} . \label{eq:alpha}
\end{equation}
For conditions typical of spiral arm regions in our Galaxy, we expect $T_* \sim T_{\rm f}$ and thus $T_{\rm f} \la T_{\rm d}$, which corresponds to $1 < \alpha \lesssim 2$.
For example, $T_{\rm f}=T_*=10$Myrs corresponds to $\alpha \approx 1.7$, which agrees well with observations \citep{Solomon+1987,Kramer+1998,Heyer+2001,RomanDuval+2010}. 

However, in a quiescent region away from spiral arms or in the outer disk, in which there is a very limited amount of dense material, $T_{\rm f}$ is expected to be larger at least by a factor of a few than in spiral arms.  
In contrast, $T_{\rm d}$ is not necessarily expected to be large even in such an environment, since the meaning of $T_{\rm d}$ is the average timescale of cloud destruction that occurs after the cloud is created, and thus, it does not necessarily depend on the growth timescale of the cloud.  
Therefore, we expect that $T_{\rm d}$ can be smaller than $T_{\rm f}$ in such an environment, which may produce $\alpha = 1+ T_{\rm f}/T_{\rm d} > 2$.  
This tendency is actually observed in Milky Way outer disk, LMC, M33, and M51 \citep{Rosolowsky2005,Wong+2011,Gratier+2012,Hughes+2010,Colombo+2014} .

The total number of molecular clouds is calculated as 
\begin{eqnarray}
  N_{\rm total} &=& \int_{M_1}^{M_2} n(M) dM 
                 =  \frac{N_0}{\alpha-1} 
                    \left[ 
                         \left( \frac{M_{\sun}}{M_1} \right)^{\alpha-1}
                       - \left( \frac{M_{\sun}}{M_2} \right)^{\alpha-1}
                    \right]    
                    \nonumber\\ 
             &\sim& \frac{N_0}{\alpha-1} 
                    \left( \frac{M_{\sun}}{M_1} \right)^{\alpha-1}, 
\end{eqnarray}
where we used $M_2 \gg M_1$ in the final estimate. 
The total number of clouds is essentially determined by the lower limit of the mass of the cloud. 
Likewise, the total mass of the molecular clouds is 
\begin{eqnarray}
  M_{\rm total} &=& \int_{M_1}^{M_2} M n(M) dM 
                 =  \frac{N_0 M_{\sun}}{2-\alpha} 
                    \left[ 
                         \left( \frac{M_2}{M_{\sun}} \right)^{2-\alpha}
                       - \left( \frac{M_1}{M_{\sun}} \right)^{2-\alpha}
                    \right]
                    \nonumber\\ 
             &\sim& \frac{N_0 M_{\sun}}{2-\alpha} 
                         \left( \frac{M_2}{M_{\sun}} \right)^{2-\alpha}, 
\end{eqnarray}
where we used $M_2 \gg M_1$ and $\alpha < 2$ in the final estimate. 
Thus, the total mass of molecular clouds is essentially determined by the upper limit of the mass of the cloud. 
Let us assume $M_{\rm total} \sim 10^9 M_{\sun}$ in the Galaxy, then our simple choice of $M_1=10^2 M_{\sun}$, $M_2=10^6 M_{\sun}$, and $\alpha=1.5$ corresponds to $N_{\rm total}\sim10^5$ and the average mass of molecular clouds is $M_{\rm ave} \equiv M_{\rm total}/N_{\rm total} \sim 10^4 M_{\sun}$. 
Note that these numbers depend on our choice of $M_1$.

\section{Summary}
In general, dense molecular clouds cannot be created in shock waves propagating in magnetized WNM without cold HI clouds. 
In this paper we identify repeated interactions of shock waves in dense ISM as a basic mechanism for creating filametary molecular clouds, which are ubiquitously observed in the nearby ISM \citep{Andre+2014}. 
This suggests an expanding-bubble-dominated picture of the formation of molecular clouds in our Galaxy, which enables us to envision an overall picture of the statistics of molecular clouds and resultant star formation.  
Together with the findings of our previous work, our conclusions are summarized as follows: 
\begin{enumerate}
\item 
Turbulent cold HI clouds embedded in WNM can be readily created in the expanding shells of HII regions or in the very late phase of supernova remnants. 
In contrast, the formation of molecular clouds in a magnetized ISM needs many compression events. 
Once low-mass molecular clouds are formed, an additional compression creates many filamentary molecular clouds. 
One compression corresponds to of order 1Myr on average in our Galaxy.
The timescale of cloud formation is a few times 10Myrs. 
\item 
Since the galactic thin disk is occupied by many bubbles, molecular clouds are formed in the overlapping regions of (old and new) bubbles.  
However, since the average lifetime of each bubble is shorter than the timescale of cloud formation, it is difficult to observationally identify the multiple bubbles that created the molecular clouds. 
\item 
The velocity dispersion of molecular clouds should originate in the expansion velocities of bubbles.  
This is estimated to be $\lesssim$10km/s and should not strongly depend on the mass of the molecular cloud. 
\item 
To describe the growth of molecular cloud mass we can temporally smooth out the evolution over timescales larger than $\sim$ 1Myr. 
The resultant mass (smoothed over time) of each molecular cloud is an almost exponentially increasing function of time. 

\item 
The destruction of a molecular cloud is mainly due to UV/FUV radiation from massive stars more massive than $20M_\sun$. 
The probability of cloud destruction is not a sensitive function of the mass of molecular clouds.  
If the shape of the initial mass function does not vary much with the mass of parent molecular clouds, cloud destruction by $30 \pm 10 M_\sun$ stars results in a star formation efficiency of order 1\%. 
This property explains the observed constancy of the gas depletion timescale ($1 \sim 2$ Gyr) of giant molecular clouds in the solar neighborhood and possibly in some external galaxies where the normalizations for the Schmidt-Kennicutt Law obtained by high-density tracers are shown to be similar.

\item 
The steady state of the evolution of the cloud mass function corresponds to a power law with exponent $-n$ in the range $1 < n \lesssim 2$ in the spiral arm regions of our Galaxy. 
However, a larger value of the exponent, such as $n > 2$, is possible in the inter-arm regions.  
\end{enumerate}

Note that the first and third conclusions have partly shown in our previous investigations \citep{InoueInutsuka2009}. 
In addition we can suggest the following implications from these conclusions: 
\begin{enumerate}
\setcounter{enumi}{6}
\item 
Star formation starts, even in small molecular clouds, once the line-mass of an internal self-gravitating filament exceeds the critical value \citep{Andre+2010}.  
Our analysis suggests that the mass of an individual molecular cloud increases roughly exponentially over $\sim 10$ Myrs.
According to the formation mechanism driven by repeated compressions, we expect that the total mass in filaments of sufficiently high line-mass increases with the number of compressional events. 
This means that the mass of star-forming dense gas increases with the mass of the molecular cloud and the star formation should accelerate over many million years. 
This conjecture may provide a clue in understanding the star formation histories found by \citet{PallaStahler2010} in seven individual molecular clouds such as Taurus-Auriga and $\rho$ Ophiuchi. 
\item 
Molecular clouds may collide over a timescale of a few times 10 Myrs, depending on the relative locations in adjacent (almost invisible) bubbles. 
Such a molecular cloud collision may result in active star formation in a giant molecular cloud \citep[e.g.,][]{Fukui+2014}. 
\end{enumerate}

These implications should be investigated in more detail by numerical simulations. 
Our radiation magnetohydrodynamics simulations of an expanding bubble due to UV and FUV photons from the massive star show that most of the material in molecular clouds become warm molecular clouds without CO molecules. 
Although we have to investigate the fate of the CO-dark gas in more detail, we expect that the total mass can be very large and may account for the dark gas indicated by various observations \cite[e.g.,][]{Grenier+2005}. 

There are many report that the Kennicutt-Schmidt correlation varies with some properties of galaxies 
\citep[e.g., ][]{Saintonge+2011,Saintonge+2012,Meidt+2013,Davis+2014}. 
In addition, a simple relation does not fit to the center of our Galaxy \citep[e.g.,][]{Longmore+2013}.  
The reasons for these deviations remain to be studied.


\begin{acknowledgements}
SI thanks Hiroshi Kobayashi and Jennifer M. Stone for useful discussions and comments. 
SI is supported by Grant-in-Aid for Scientific Research (23244027,23103005)
\end{acknowledgements}


\end{document}